# Can gamification help in software testing education? Findings from an empirical study

[1]Raquel Blanco, [2]Manuel Trinidad, [1]María José Suárez-Cabal, [2]Alejandro Calderón, [2]Mercedes Ruiz, [1]Javier Tuya

[1]Software Engineering Research Group, University of Oviedo, Department of Computer Science,

Gijón, Spain

{rblanco, cabal, tuya}@uniovi.es

[2]Software Process Improvement and Formal Methods Research Group, University of Cadiz,

Department of Computer Science and Engineering, Cádiz, Spain

{manuel.trinidad, alejandro.calderon, mercedes.ruiz}@uca.es

**Abstract**

Software testing is an essential knowledge area required by industry for software engineers. However, software engineering students often consider testing less appealing than designing or coding. Consequently, it is difficult to engage students to create effective tests. To encourage students, we explored the use of gamification and investigated whether this technique can help to improve the engagement and performance of software testing students. We conducted a controlled experiment to compare the engagement and performance of two groups of students that took an undergraduate software testing course in different academic years. The experimental group is formed by 135 students from the gamified course whereas the control group is formed by 100 students from the non-gamified course. The data collected were statistically analyzed to answer the research questions of this study. The results show that the students that participated in the gamification experience were more engaged and achieved a better performance. As an additional finding, the analysis of the results reveals that a key aspect to succeed is the gamification experience design. It is important to distribute the motivating stimulus provided by the gamification throughout the whole experience to engage students until the end. Given these results, we plan to readjust the gamification experience design to increase student engagement in the last stage of the experience, as well as to conduct a longitudinal study to evaluate the effects of gamification.

## 1 Introduction

In the Information Technology era, software is present in almost every aspect of our lives. It is crucial that software products work correctly and as intended, because their malfunction could produce financial loss to both final users and software companies, loss of prestige to software companies, exposure of private data, or even tragic consequences in human life. In order to prevent these negative effects, it is essential to test properly and thoroughly the software products with the aim of detecting and solving their defects (also called faults or bugs) and failures before releasing them to final users. For that purpose, industry requires well-prepared testers and education has a responsibility for that issue. However, there is a gap between industry and education regarding the testing knowledge needed in the industry and what is taught in higher education institutions. On the one hand, software testing is not taught with enough depth in the software engineering education (Fraser et al., 2019; Jesus et al., 2020) and it is often embedded in other software engineering courses instead of being a dedicated course (Silvis-Cividjian, 2021). On the other hand, it is difficult to engage students as they tend to find software testing a boring and destructive task, and they prefer the design and implementation tasks (Fraser et al., 2020). As a result, students do not acquire the practice that the industry requires and they prefer to work in other software engineering areas rather than in software testing; as stated in (Silvis-Cividjian, 2021), only up to 20% of students consider a career in software testing.

To deal with the problem of motivating students, we have explored the use of gamification in a software testing undergraduate course. Gamification has been adopted in the last decade as a means to engage students in different educational contexts (Bodnar et al., 2016; Dichev & Dicheva, 2017; Milosz & Milosz, 2020; Subhash & Cudney, 2018; Vos et al., 2020), and has given promising results in software engineering and software testing education (Alhammad & Moreno, 2018; Fraser, 2017; Garousi et al., 2020; Jesus et al., 2018; Kosa et al., 2016; Pedreira et al., 2015; Souza et al., 2018).

In this paper, we describe and evaluate the gamification experience carried out in seminar classes of a software testing course, where students had to create effective test cases for several faulty programs in order to detect their defects. We hypothesized that gamification would positively impact on student engagement and performance. The students were involved in a narrative-based gamification experience where they took part in a legendary Olympic race for immortality, supported by a gamification tool called GoRace. The effectiveness of their test cases determined their progress in the race. This effectiveness is provided by a tool called SQLTest, which is integrated with GoRace. To test our hypotheses, we defined a set of metrics to measure the student engagement and performance and statistically compared the results achieved by the experimental and the control groups in order to answer the following research questions:

- RQ1. Is the engagement of the software testing students who carry out gamified activities higher than the ones who carry them out in a non-gamified environment?
  To address this research question, we defined a set of metrics related to the student's participation in the academic activities.
- RQ2. Is the performance of the software testing students who carry out gamified activities higher than the ones who carry them out in a non-gamified environment?

To answer this research question, we defined a set of metrics related to the effectiveness of the test cases created.

The main contributions of this work are as follows:

- The design of a gamification experience for an undergraduate software testing course. This experience was carried out along a complete academic semester (15 weeks).
- An analysis of the impact of gamification on improving the student engagement and their performance in the creation of effective test cases. The results indicate that gamification has a positive impact on both the student engagement and performance; in addition, they also reveal that it is crucial to maintain the stimulus during the whole experience to ensure success.

The rest of this paper is organized as follows: Section 2 provides a background on gamification and software testing education. Section 3 presents the course involved in the gamification experience, along with the participants and the materials used in the experience. This section also describes the procedure followed in the experience and the metrics defined. Section 4 presents the results of the statistical analysis carried out to test the hypotheses that allow us to answer the research questions. These results are discussed in Section 5. Section 6 lists the threats to the validity of the findings discussed. Finally, we draw our conclusions and describe the implications of our findings in Section 7.

## 2    Gamification and software testing education

### 2.1    Software testing

Software testing is a set of activities involved in software development conducted to determine whether a software product satisfies the specified requirements and fits the user needs, as well as to detect failures and defects (International Software Testing Qualification Board (ISTQB), 2020). One of these activities encompasses the design and implementation of sets of test cases, called test suites. A test case consists of the program input (test input) and the expected output that should be obtained. The execution of a test case against the program under test allows the tester to observe whether there is any deviation between the output obtained and the expected output. In that case, a failure is found. A failure is caused by the existence of defects in the program under test.

Software testing is crucial for evaluating and assuring the quality and for reducing the risk of failure when a product is released. The percentage of the software development budget allocated to software testing in 2020 was 22% (Capgemini, 2021). Despite that, the total cost of poor software quality continues to trend upward. As stated in (Krasner, 2021), from the total cost of poor software quality in the US in 2020 ($2.08 trillion), $1.56 trillion was due to software failures, which has grown 22% over the last 2 years. This cost could have been reduced with comprehensive testing.

Testing all the possible input and output combinations of a program is impractical, and often impossible (Myers et al., 2012). For this reason, it is essential to design and implement test suites that are effective in order to reduce the impact of software defects and failures and the cost of software testing. The effectiveness of a test suite is its ability to find defects: the more defects it is capable of finding, the more effective it is.

Despite the importance of software testing, it is frequently neglected in computer science education (Jesus et al., 2020), where the amount of time spent studying software testing is significantly less than that spent on other software development activities (Sherif et al., 2020; Vos et al., 2020; Zivkovic & Zivkovic, 2021). Dedicated courses on software testing are not very common (Silvis-Cividjian, 2021) and, even when the curricula include them, much more effort has to be made in order to provide the students with practical problems (Zivkovic & Zivkovic, 2021). Besides, engaging students is challenging for several reasons. On the one hand, testing is perceived as a destructive task (Myers et al., 2012), while software design and implementation are considered more creative activities. Thus, students are less interested in software testing (Deak et al., 2016; Garousi et al., 2020; Vos et al., 2020). Instead of perceiving the benefits of finding defects to improve quality, students do not feel satisfaction when the defects of their own programs are exposed, as they signal their code is not correct. Therefore, students are not motivated to find these defects (Garousi et al., 2020). On the other hand, software testing education tends to be more theoretical than practical, and very frequently students describe it as boring (Fraser et al., 2019; Garousi et al., 2020), especially when they find few to no defects. Moreover, the lack of real-life testing scenarios creates the sense of doing something repetitive and irrelevant (Isomöttönen & Lappalainen, 2012). In addition, if the programs to be tested contain a few meaningful defects, students tend to perceive testing as a tedious and unsatisfying task (Silvis-Cividjian, 2021). Overcoming these problems is critical to improving the students' testing skills.

## 2.2 Gamification

In order to engage software testing students, several learning environments and educational approaches have been devised, such as web-based tutorials (Elbaum et al., 2007), flipped classrooms (Elgrably & Oliveira, 2022), problem-based learning (Andrade et al., 2019), serious games (Valle et al., 2017), as well as gamification (Jesus et al., 2018), which is explored in this work.

Several definitions have been provided in the literature for the term gamification, such as "the use of game elements and game design techniques in non-game contexts" (Werbach & Hunter, 2012), "the phenomenon of creating gameful experiences" (Hamari et al., 2014), "the use of typical elements of games in contexts outside the game environment" (Deterding et al., 2011), or "the use of game elements in non-gaming context to boost engagement between humans and computers and resolve issues with high quality modern electronic applications" (Khaleel et al., 2016). They all agree that gamification is not the creation of a fully-fledged game, but consists of applying lessons from the game domain to increase commitment and motivation in non-game situations (Calderón et al., 2018).

In recent years, gamification has attracted the attention from both practitioners and researchers as a way to achieve a range of emotional, cognitive, and social benefits, and guide human behavior for inducing innovation, productivity, or engagement (Sardi et al., 2017) in different contexts, such as employee performance, customer engagement and social loyalty, and in a diversity of domains, including marketing, human resources, healthcare, education, environmental protection and wellbeing (Dichev & Dicheva, 2017).

Software engineering has also explored the strengths and weaknesses of applying gamification for the learning of the software engineering processes. During the last decade, several works have applied gamification in order to improve student's engagement, performance and social skills, as well as to encourage the use of software engineering best practices (Alhammad & Moreno, 2018; Garcia et al., 2020). The results obtained are promising, but the research on gamification in software engineering education is still at an early stage. Thus, further research and more empirical studies are needed for analyzing whether gamification is a useful and effective technique in this context (Alhammad & Moreno, 2018; Kosa et al., 2016; Pedreira et al., 2015; Souza et al., 2018).

## 2.3 Gamification in software testing education

Previous works that applied gamification in software engineering education are mainly focused on the software construction and software engineering process areas of the SWEBOK guide (Bourque & Fairley, 2014) (Alhammad & Moreno, 2018; Pedreira et al., 2015; Souza et al., 2018). Software testing has also attracted the researchers' interest and it is a promising area for applying gamification. Some works are focused on applying gamification to expose students of introductory computer science courses to software testing, such as (Bell et al., 2011; Sheth et al., 2015, 2013). Other works apply gamification to engage students in the learning of Agile test practices (Elgrably & Oliveira, 2018; Lőrincz et al., 2021), unit testing (Marabesi & Silveira, 2019), Graphical User Interface testing (Cacciotto et al., 2021; Garaccione et al., 2022), testing tools (Clarke et al., 2017; Fu & Clarke, 2016) or test incident reporting (Dal Sasso et al., 2017).

Most of the works use gamification to engage students in learning several testing techniques, such as the code review process (Dal Sasso et al., 2017; Khandelwal et al., 2017), exploratory testing (Costa & Oliveira, 2019, 2020; Lőrincz et al., 2021), statement coverage (Clegg et al., 2017; Sherif et al., 2020), loop coverage (Clegg et al., 2017), control flow testing (Buckley & Clarke, 2018; Clarke et al., 2019, 2022, 2020), dataflow testing (Buckley & Clarke, 2018; Clarke et al., 2019, 2022, 2020; Clegg et al., 2017), equivalence partitioning (Buckley & Clarke, 2018; Clarke et al., 2019, 2022, 2020; Jesus et al., 2020), boundary value analysis (Buckley & Clarke, 2018; Clarke et al., 2019, 2022, 2020; Clegg et al., 2017; Jesus et al., 2020), state-based testing (Buckley & Clarke, 2018; Clarke et al., 2019, 2022, 2020) or mutation testing (Rojas & Fraser, 2016). Other works use gamification to motivate students to create test cases for findings bugs, such as (Fraser et al., 2019, 2020; Silvis-Cividjian, 2021), which are the closest works to ours. (Silvis-Cividjian, 2021) presents an approach that uses the platform VU-BugZoo, which contains embedded buggy code. Students have to design test strategies and create test cases to find the defects. (Fraser et al., 2019, 2020) use the game Code Defenders to engage students to test Java classes. Students play as attackers, who modify the source code to introduce artificial defects, or defenders, who implement test cases in Junit that reveal the existence of those defects.

Table 1 summarizes the aforementioned works and compares them with ours. First, works are classified according to whether they report a practical experience. When a practical experience is reported, its scope, duration and number of participants are indicated. Table 1 also indicates whether a comparison with a non-gamified experience (control group) is made, the number of participants in the control group and whether a statistical analysis is reported. In general, the extent of practical experiences reported is small: the number of participants is not very large, which makes it difficult to extrapolate the results, and/or the experiences are of limited time length or are applied to individual assignments, so it is difficult to analyze the long-term impact of gamification. On the other hand, these gamification experiences are not compared against non-gamified ones in most of the works. Besides, statistical analysis is not commonly reported. Our work shares the aims of engaging students and improving their performance with the foregoing works, but unlike them, we present a long gamification experience that lasted a whole academic semester (15 weeks), where 135 students were involved and the rewards were given at the end of the experience. In addition, we conducted a controlled experiment that qualitatively compares the gamification experience against a non-gamified one and we carried out a statistical analysis to test the hypotheses stated.

| Work | Practical experience (yes/no) | Scope | Duration | Participants | Comparison with control group (yes/no) | Participants on control group | Statistical analysis (yes/no) |
|---|---|---|---|---|---|---|---|
| (Bell et al., 2011) | No | - | - | - | - | - | - |
| (Buckley & Clarke, 2018) | Yes | Assignments | 1 semester | 18 | Yes | 18 | No |
| (Cacciotto et al., 2021) | No | - | - | - | - | - | - |
| (Clarke et al., 2017) | Yes | Assignments | 13 weeks | 3 groups: 24, 33 and 27 | No | - | No |
| (Clarke et al., 2019) | Yes | Assignments | 90 minutes | 19 | Yes | 20 | Yes |
| (Clarke et al., 2020) | Yes | Assignments | 155 minutes | 2 groups: 11 and 15 | No | - | No |
| (Clarke et al., 2022) | Yes | Assignments | 90 minutes | 62 | Yes | 60 | Yes |
| (Clegg et al., 2017) | No | - | - | - | - | - | - |
| (Costa & Oliveira, 2019) | No | - | - | - | - | - | - |
| (Costa & Oliveira, 2020) | Yes | Classes | Case study 1: 7 days Case study 2: 8 days | Case study 1: 3 Case study 2: 6 | No | - | No |
| (Dal Sasso et al., 2017) | No | - | - | - | - | - | - |
| (Elgrably & Oliveira, 2018) | Yes | Course | Case study 1: 6 sessions Case study 2: 7 sessions | Case study 1: 14 Case study 2: 20 | No | - | No |
| (Fraser et al., 2019) | Yes | Lab activities gamified separately | 70 minutes per session (12 sessions) | 123 | No | - | Yes |
| (Fraser et al., 2020) | No | - | - | - | - | - | - |
| (Fu & Clarke, 2016) | Yes | Classes | 1 session | 3 groups: 33, 15 and 13 | No | - | No |
| (Garaccione et al., 2022) | No | - | - | - | - | - | - |
| (Jesus et al., 2020) | Yes | Course | 4 hours | 2 groups: 11 and 16 | Yes | 22 | Yes |
| (Khandelwal et al., 2017) | Yes | Assignment | 15 days | 3 groups: 37, 38 and 36 | Yes | 2 groups: 34 and 36 | No |
| (Lőrincz et al., 2021) | Yes | Lecture, seminar and lab activities (lectures, seminars and labs are gamified separately) | Lectures: 1 session Labs: 10 sessions Seminars: 1 session | n/a | In labs | n/a | No |
| (Marabesi & Silveira, 2019) | No | - | - | - | - | - | - |
| (Rojas & Fraser, 2016) | No | - | - | - | - | - | - |
| (Sherif et al., 2020) | Yes | Assignment | Short (according to authors) | 10 | Yes | 10 | Yes |
| (Sheth et al., 2015, 2013) | Yes | Assignments gamified separately | 2 weeks per assignment (3 assignments) | 124 | No | - | No |
| (Silvis-Cividjian, 2021) | No | - | - | - | - | - | - |
| Our work | Yes | Assignments | 15 weeks | 135 | Yes | 100 | Yes |

Table 1. Related works on gamification in software testing education.

# 3 Experiment design

This section presents the design of the controlled experiment reported in this paper. Section 3.1 and Section 3.2 describe the course in which this experiment took place and the participating students, respectively. The materials used by the students in the experimental and control groups are described in Section 3.3 and the activities they carried out are explained in Section 3.4. Finally, the metrics defined to analyze the results of the experiment are presented in Section 3.5.

## 3.1 Course design

The experiment takes place in the course titled "Software Verification and Validation" of the $4^{th}$ year of the Software Engineering degree of the University of Oviedo (Spain). This is a one-semester course (15 weeks) addressing the topics related to the testing process and techniques required to perform effective software tests. It consists of the following types of components:

- Lectures, in which the professor presents the theoretical contents of the course and provides explanation, examples and practical recommendations. Lecture classes deal with the evaluation of software quality, from the product point of view, and are mainly focused on software testing. The software testing process is addressed, as well as the testing techniques and strategies to be used to create effective test suites. The assessment of the theoretical contents of the course is based on an exam, which is worth 35% of the final grade.
- Seminars, in which the students address the challenge of creating effective test suites following a continuous improvement approach. For that purpose, the students select and apply testing techniques that were introduced previously in the lecture classes. The students work on a sequence of four exercises, evenly distributed over the whole semester, with an increasing level of difficulty. For each exercise, the students work in teams of 3 or 4 members to create test suites for testing a program with injected defects provided by the professor. After that, they work individually to improve continuously the test suite effectiveness, in order to find as many defects as possible. Table 2 presents a brief description of the program to be tested in each exercise, as well as the number of injected defects and some examples of these defects. Seminar contents and skills are assessed based on: for each exercise, the quality of a test design report provided by the students and the effectiveness of the test suite submitted determined by the SQLTest tool, which is worth 20% of the final grade. This report is submitted before the effectiveness measurement and contains: the test design, an explanation of the testing techniques used and the test cases that constitute the test suite, along with the traceability between the test design and the test cases. The quality of the test design report is graded on the basis of the structure of the test design and the test cases, the correct use of the testing techniques and the degree of traceability established.

- Laboratories, in which the students put into practice the contents and skills learned in the lectures and seminars to conduct a software testing project for a real-life application provided by the professors. This project is performed individually and involves the following test processes: test design and implementation, test environment set-up and maintenance, test execution and test incident reporting. Laboratories are assessed based on the continuous assessment of each student's work and progress, which is worth 15% of the final grade, and the quality of the assignments submitted, which is worth 30% of the final grade. These assignments are the test design, the test suite created, along with the traceability between the test design and the test suite, the result of the test execution and the test incident reports.

Our experiment was conducted in the seminars, which consist of weekly one-academic-hour sessions.

| Exercise | Program | Description | Number of injected defects | Examples of injected defect |
|---|---|---|---|---|
| 1 | Triangle classification | It finds the type of triangle based on the lengths of its sides. | 16 | - Length side 2 + length side 3 < length side 1: it is classified as a triangle. (For example, the input *4 2 1* is classified as a triangle).<br>- Only two equal sides: it is classified as equilateral. (For example, the input *2 2 3* is classified as equilateral).<br>- 3 equal letters introduced as the sides of the triangle: it is classified as an equilateral triangle. (For example, the input *a a a* is classified as an equilateral triangle). |
| 2 | On-line shop | It shows the products of a shop, according to several search criteria that have to be fulfilled simultaneously. | 20 | - Products that do not fulfill the criterion "products whose description starts with ___" are shown when the description contains the value searched. (For example, when the criterion "products whose description starts with big" is established, a product whose description is "the book tells the story of Abigail" is shown).<br>- Products whose price is equal to the value specified in the criterion "products whose minimum price is ___" are not shown. (For example, when the criterion "products whose minimum price is 10" is established, a product whose price is 10 is not shown).<br>- Criterion "products whose description contains ___" distinguishes upper case text from lower case text. (For example, when the criterion "products whose description contains iron" is established, a product whose description is "the superhero Iron Man fights against his enemies" is not shown). |
| 3 | Spanish VAT return | It generates a new VAT return, based on the information of previous VAT returns stored in a | 15 | - Very small differences (cents of euro) between output VAT and input VAT are not considered to calculate the VAT payable. (For example, when the difference |

| | | database and the taxable transactions of the corresponding period. | | between the output VAT and the input VAT is 0.01€, the program uses the value 0 to calculate the VAT payable).<br>- The reduction of the VAT payable stops when a previous period has a difference between output VAT and input VAT equal to 0. (For example, the database stores the VAT return of the first three trimesters of 2022 and the differences between the output VAT and the inputs VAT are -100€, 0€ and -200€, respectively. In the fourth trimester, the difference between the output VAT and the input VAT is 350, and when the program calculates the VAT payable, it stops the reduction in the second semester, instead of using the three trimesters).<br>- Non-allowed periods to make VAT adjustments are used. (For example, the database stores the VAT return of the four trimesters of the years 2021 and 2022. These VAT returns have a negative difference between the output VAT and the input VAT. To simplify, we consider that we can only use one year to reduce the VAT payable. When the program calculates the VAT payable for the first trimester of 2023, it uses the four trimesters of 2022 and the last trimester of 2021, instead of using only the four trimesters of 2022). |
|---|---|---|---|---|
| 4 | Lab reservation clash | It checks whether the lab reservation a professor wants to make clashes with the existing reservations of other courses of the same university degree and academic year. | 11 | - It is not checked whether the university degree of the course involved in the new reservation is the same as the university degree of the reservations already made in order to detect a clash. (For example, a professor wants to make a lab reservation for the course "Software Verification and Validation" of the 4[th] year of the Software Engineering degree. The same day and at the same time, there is a lab reservation for the course "Web Technologies" of the 4[th] year of the IT Engineering degree. The program does not check the university degree and it determines that there is a clash, when in fact there is not).<br>- The reservations already made for the course involved in the new reservation produce a clash, but they do not have to. (For example, a professor wants to make a lab reservation for the course "Software Verification and Validation" of the 4[th] year of the Software Engineering degree. The same day and at the same time, there is a lab reservation for the same course. The program does not exclude the lab reservations already made for the |

| | | | | | same course and determines that there is a clash, when in fact there is not. Note that a course can have several professors that can conduct the lab classes of different groups simultaneously.)<br>- A reservation already made that finishes the same day that the new reservation starts does not produce a clash. (For example, a professor wants to make a lab reservation from 17:00 to 18:00, starting on January 16th, 2023. There is a reservation for another course of the same degree and year from 17:00 to 18:00 that finishes on January 16th, 2023. The program does not compare both dates correctly and determines that there is not a clash, but actually there is.) |

Table 2. Programs to be tested in the exercises of seminars: number of the exercise, program name, brief description of the program, number of injected defects and examples of some of the injected defects.

### 3.2 Participants

The controlled experiment consisted of one experimental group and one control group. The experimental group is formed by the students that voluntarily got involved in the gamification experience performed during the seminars in the academic year 2020-2021. The number of participating students was 135 (98% students in the 18-28 age range and 2% in the 28-38 age range), which corresponds to 94% of the students enrolled in the course.

This study obtained the approval from the Responsible Research and Innovation Subcommittee of the Research Ethics Committee of the University of Oviedo and is conformed to the ethical principles and Spanish legislation. The main ethical aspects of this study involved the participants' informed consent and the personal data protection. All the participants were informed of the experiment, all the data to be collected and their treatment. They gave their consent before entering the research and answering a demographic questionnaire. The participants were informed that their participation in the experiment was voluntary and the decision to participate or not participate, or to discontinue participation, will not result in any consequences, academic or otherwise.

The control group is formed by the 100 students who studied the same course in the academic year 2019-2020, where the seminar activities were not gamified. The age range is similar to the experimental group, but precise age data are not available because the control group did not answer a demographic questionnaire.

All seminar sessions, in both the experimental and control groups, were conducted by the same professor, who has experience in software testing teaching and researching.

## 3.3 Materials

The materials used in this experiment are the programs to be tested, provided to the students as learning materials, and the software applications SQLTest and GoRace, which have been integrated to support the gamification experience and are used by the students to carry out the learning activities. The following subsections provide further information about them.

### 3.3.1 Programs to be tested

As mentioned before, the experiment was conducted in the seminar classes where the students work on a sequence of four exercises (see Table 2) to apply their knowledge and develop their skills as software testers. A critical resource for these exercises is, therefore, the program that each testing exercise is focused on.

For each program, the following materials are created and provided by the professors:

- Program specification. It describes the program functionality. The students are provided with the program specification, as prepared by the professors, and they used it as the body of knowledge from which the test suites are created. Table 2 of Section 3.1 provides a brief functional description of each of these programs.
- Defects. For each program, the professors decide and create manually a set of significant defects tailored to the particular program functionality, which usually appear during the software development. Each defect produces a different failure. Initially, the students do not know these defects. Table 2 of Section 3.1 indicates the number of defects created for each program, as well as some examples of these defects.
- Program implementations. For each program, the professors create several independent implementations, called *versions*, for the corresponding program specification. One of these versions implements the program specification correctly and it is called the *original version*. The remaining versions are faulty ones, called *mutants*. Each mutant is an originally correct implementation modified to contain only one of the defects injected manually by the professors. Therefore, each program has as many mutants as tailored defects which were created for that program. In addition, the source code of the implementations is not available to students.

### 3.3.2 SQLTest

SQLTest is a software tool developed by the Software Engineering Research Group of the University of Oviedo. SQLTest allows the students to execute their test suites and evaluate their effectiveness for the program implementations that have been previously loaded by the professors. SQLTest was initially created to test SQL queries, hence its name. Later, it was expanded to test programs in general, but it was decided to keep its original name.

For each exercise, SQLTest embeds all the implementations of the program to be tested, that is, the original version and the mutants. When a student submits their test suite for a particular exercise, SQLTest executes this test suite internally against the original version and all the mutants and compares the outputs obtained by each execution. If the output of the original version is different to the output of a mutant, that means that the defect injected in such a mutant has been detected by the test suite provided by the student.

SQLTest determines the test suite effectiveness as the percentage of detected defects over the total number of defects to be detected, that is, the number of mutants. After that, SQLTest gives feedback to students, indicating the test suite effectiveness, as well as the description of the defects that have not been detected yet if that is the case.

### 3.3.3 GoRace

GoRace is a multi-context and narrative-based gamification tool developed by the Software Process Improvement and Formal Methods Research Group of the University of Cadiz. GoRace allows the user to automatically create a tailored web solution to gamify the use of any third-party tool with which it can be easily integrated through a REST API.

In a GoRace experience, the participants are immersed in a virtual world based on Greek mythology where they take part in an Olympic race decreed by Zeus to commemorate his victory over his father Cronus. The prize for those who reach the finish line of this legendary race is immortality. To provide a meaningful gamified experience for the different types of existing players, GoRace does not only implement the well-known and widely used game elements such as points and leaderboards, but a comprehensive range of game elements such as: a) game dynamics, such as sense of progress, sense of competence, interaction, narrative and socialization, b) game mechanics, such as challenges, effort, participation, strategies and encouragement, and c) game components, such as results, evolution, classification, points and gifts). In total, GoRace implements 131 different game elements out of the 229 game elements identified by Peixoto and Silva in their systematic literature review (Peixoto & Silva, 2017). More information about the game elements implemented in GoRace can be found in (Trinidad et al., 2021).

An Olympic race in GoRace takes place between two places that are decided when creating a gamified experience. In the case of the experiment described in this paper, the School of Computer Engineering in Oviedo, where the students attend their classes, was selected as the starting point, and the beach of Gijón, where the students enjoy their leisure time, was selected to be the finish line. The participants in the race need to cover the distance between those two places. Such distance is measured in distance units. Figure 1 shows a screenshot of GoRace displaying the map of the Olympic race. As it can be seen, a blue line shows the route of the race and over it the different avatars of the participants are placed, according to their position in the race.

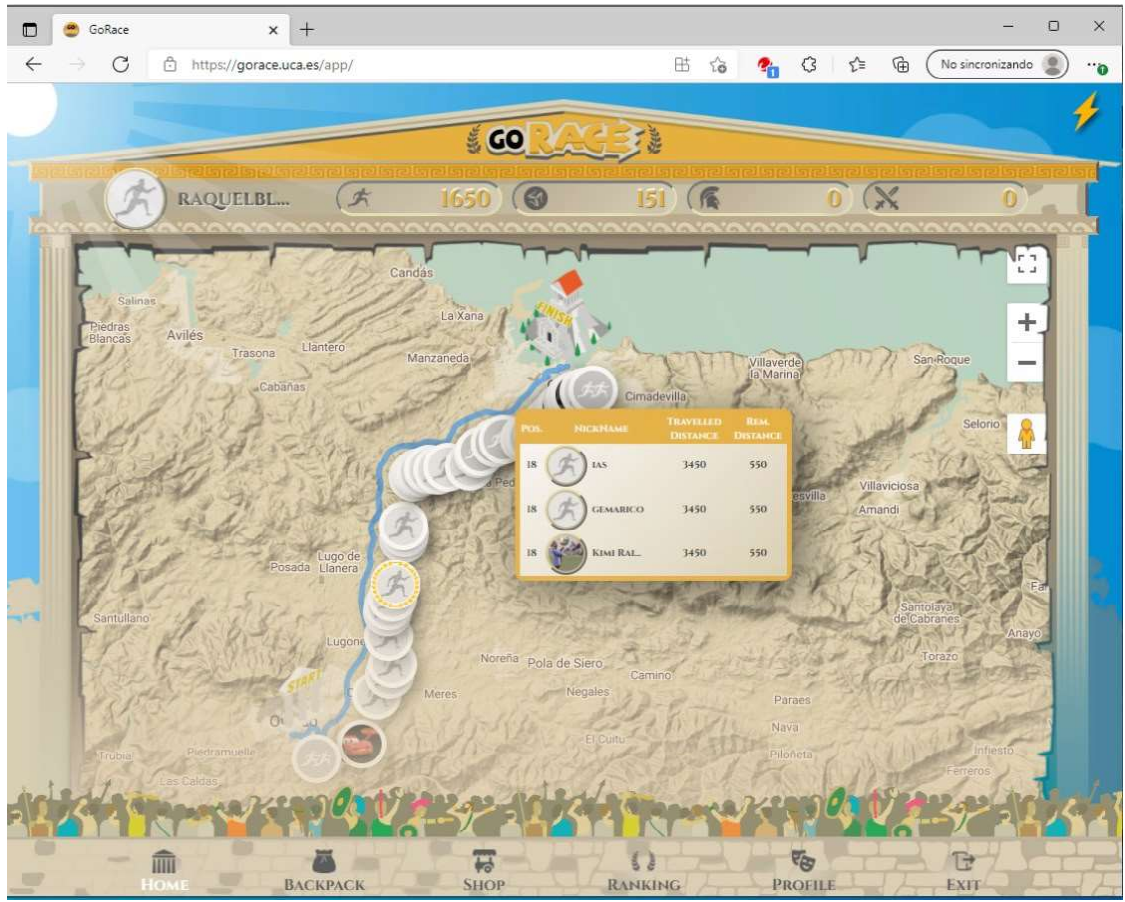

Figure 1. Map of the Olympic race in GoRace that shows the student position (circle with the border highlighted in yellow) and the positions of the competitors.

During the race, the participants can obtain divine points, a type of virtual currency, that they can use to purchase different powerful goods called Relics. Once purchased, the participant needs to decide when is the most suitable moment to make use of their Relics' powers. The Relics have powers with positive or negative consequences, such as different distance increments for advancing a specific number of distance units in the race, attacks for decreasing the distance units achieved by other players or protections for neutralizing an attack or reflecting it back onto the attacker.

### 3.3.4  Integration of SQLTest and GoRace

In the gamification strategy implemented in GoRace for this experiment, both the distance units traveled in the race and the divine points obtained by the students are based on the assessment of the effectiveness of their test suites determined by SQLTest: the higher the effectiveness, the higher the distance units and the divine points the student is rewarded with. As both distance units and divine points are positive rewards to motivate the students to work on the seminars, negative rewards have not been used in the experiment.

The integration of SQLTest and GoRace was done by using the API provided by GoRace. Through that API, SQLTest sends to GoRace the data containing the identification of the exercise and the student, along with the effectiveness of the test suite executed and the timestamp of this execution. Then, GoRace calculates the distance units and the divine points the student will be awarded with as a function of their effectiveness reported by SQLTest and updates the information shown on the application screens, such as the map of Figure 1. The interaction between SQLTest and GoRace is depicted in Figure 2.

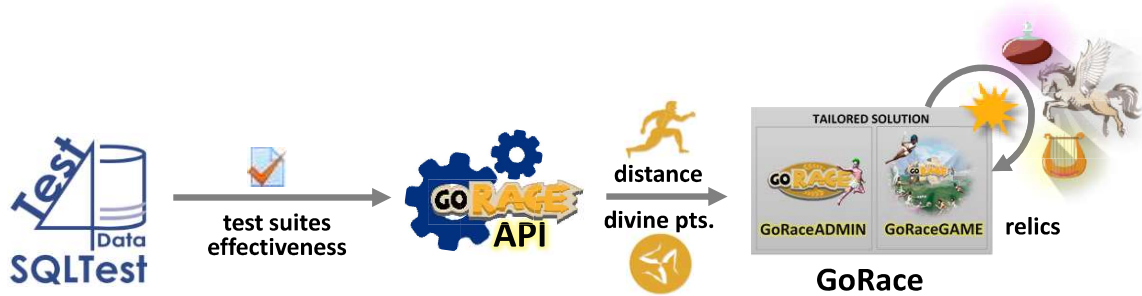

Figure 2. SQLTest-GoRace interaction

### 3.4 Procedure

In our experiment, the Olympic race is made of the four exercises presented in Section 3.1 (see Table 2), carried out in sequence. In the first seminar session, the gamification experience and the Olympic race were introduced to the students, as well as the tools SQLTest and GoRace. The instructions to be followed to carry out the exercises were explained too. The students were informed of the experiment, the treatment of the data to be collected and their voluntary participation. After that, they answered the demographic questionnaire.

The remaining seminar sessions and homework implemented the Olympic race, which covered 4000 distance units, during 14 weeks (Table 3 shows the duration of each exercise in hours). The race started in the first exercise and the subsequent ones allowed the students to progress in the race. The test suite effectiveness achieved in every exercise gave at most 1000 distance units and at most 90 divine points. After finalizing the fourth exercise, the race also finished. Figure 3 depicts the procedure followed in seminars and the activities involved in each exercise.

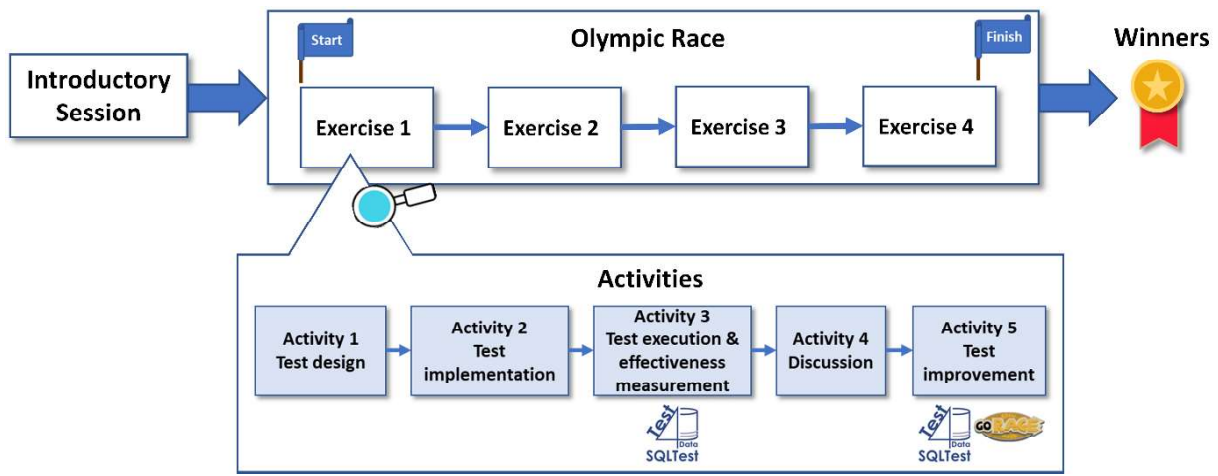

Figure 3. Procedure of seminar classes

| Exercise | Program to be tested | Duration (hours) | |
|---|---|---|---|
| | | Face-to-face classes | Homework (estimated) |
| 1 | Triangle classification | 3 | 4 |
| 2 | On-line shop | 3 | 12 |
| 3 | Spanish VAT return | 3 | 12 |
| 4 | Lab reservation clash | 3 | 12 |

Table 3. Duration in hours of each exercise: number of the exercise, name of the program to be tested in the exercise, number of hours spent in face-to-face classes and number of estimated hours that the students spent on homework.

Each exercise is composed of the following activities:

1. Test design: (one class session) the professor introduces the specification of the program to be tested, and after that the students design the tests, applying the testing techniques explained in the lecture classes. The students work in teams in this activity.

2. Test implementation: (one class session) the teams create the test cases that constitute their test suites, applying the strategies explained in the lecture classes to determine the test input of each test case.

3. Test execution and effectiveness measurement: (2 days, homework) each student individually provides SQLTest with the test suite and executes it to measure the effectiveness achieved. SQLTest shows the test suite effectiveness to the student, along with the description of the defects that have not been detected yet. This information constitutes valuable feedback that can be used to improve the test suite.

4. Discussion: (one class session) the professor and the students present several alternatives for the test design and implementation of this exercise and carry out a discussion about the effectiveness of these alternatives.

5. Test improvement: (two weeks, homework) the students, individually, use the feedback given in the discussion activity and the feedback provided by SQLTest to improve their test suite (they can add, remove and modify test cases of their test suites in SQLTest). Whenever the test suite has been modified, the student can execute it in SQLTest to measure the new effectiveness achieved and receive new feedback. After obtaining the results of each execution, SQLTest sends automatically the effectiveness to GoRace in order to update the distance units traveled in the Olympic race and the divine points. The students use GoRace to know the distance units and divine points they have reached, to check the race ranking, to buy relics that give them some benefits and to interact with other players (attacking other players or protecting themselves).

The ability to apply testing techniques to design the tests that give rise to effective test suites, as well as the ability to determine the test inputs, are crucial to obtain as many distance units and divine points as possible at the beginning of the test improvement activity of each exercise. During the test improvement activity, the abilities to understand the defects and determine the test inputs give rise to the additional distance units and divine points the students accumulate in each exercise.

The exercises were evaluated by the professor and the students received their grade on a scale from 0 to 10. Then, the 30% of the students that achieved the higher positions in the race ranking were rewarded with additional points, from 0.5 to 1.5 points.

The procedure followed by the students in the control group was similar to the procedure described above. The first seminar session addressed the same issues, except those related to the experiment and the demographic questionnaire. During the remaining 14 weeks, they worked on the same sequence of four exercises, carried out the same activities that form each exercise and used the same materials, except GoRace.

### 3.5 Metrics

The main objective of our gamification experience is to increase the engagement of the students in the test improvement activity, so that they became more motivated to use SQLTest to improve their test suites. Related to this objective, this experience also tries to increase the performance of the students in the creation of test suites.

To measure the engagement of the students, we have considered the proposal of (Fredricks et al., 2004), which divides the engagement concept into three facets (behavioral, emotional and cognitive). In particular we have considered the behavioral engagement facet, which is based on the idea of student's participation and involvement in academic and extracurricular activities. For the purpose of this study, we have defined four metrics that measure the student's academic participation in terms of their interaction with SQLTest (the first two metrics) and their interaction with the exercises (the last two metrics):

- *Number of Test Suite Executions*: it is the number of times that a student carries out the execution of a test suite in the test improvement activity of each exercise, with the aim of determining its effectiveness.
- *Active Time*: it is the number of days that a student interacts with SQLTest executing at least one test suite in the test improvement activity of each exercise. This metric measures the number of days where a student is actively interacting with SQLTest and complements the previous one. Thus, we can measure the interaction degree achieved by the students during the days they were actively working on the test improvement activity.
- *Participation Rate:* it is the proportion of students that work in all activities of each exercise, including the test improvement activity.
- *Dropout Rate*: it is the proportion of students that do not work on each exercise and abandon the rest of seminar exercises in that moment, over the number of students that have worked on the previous exercise.

To measure the performance of students, we have used two metrics related to the test suite defect finding ability, which indicate the quality of the test suite:

- *Effectiveness*: It is the percentage of defects detected by the test suite of a student over the total number of defects to be detected (that is, the mutation score).
- *Effectiveness Increase*: It is the relative effectiveness increment of the test suite of a student in the test improvement activity of each exercise over their particular margin of improvement.

For each metric, we collected data of each individual exercise for the experimental and control groups, in order to analyze each one separately. We also accumulated the data of the four exercises to analyze the experience as a whole.

## 4 Results

This section presents the experiment results and answers the research questions introduced in Section 1. The data collected to carry out the analysis are available at *http://dx.doi.org/10.17811/ruo_datasets.64866*.

### 4.1 RQ1: Student engagement

RQ1: Is the engagement of the software testing students who carry out gamified activities higher than the ones who carry out them in a non-gamified environment?

To answer the research question RQ1 we state the null hypothesis of no difference in the level of engagement of the software testing students who carry out gamified activities in comparison with the students who carry them out in a non-gamified environment. To test this hypothesis, we analyze the metrics *Number of Test Suite Executions, Active Time, Participation Rate* and *Dropout Rate*. We carried out a statistical analysis with α=0.05 to check whether the metrics in the experimental group are significantly different from the ones obtained in the control group, considering the Olympic race (the whole sequence of four exercises), as well as each exercise individually. First, the Kolmogorov-Smirnov test was performed to check the normality of the distributions. A very small p-value (<0.001) was obtained in each test, so we cannot assume that they are normally distributed. Therefore, the Mann-Whitney U test for independent samples was applied to verify the null hypothesis of median equality. Table 4 shows the results for the four metrics: mean and median for the control and experimental groups, p-value and U obtained by the Mann-Whitney U test. To measure the effect size, we have used $r$ ($r<0.1$ no effect, $0.1 \leq r < 0.3$ small, $0.3 \leq r < 0.5$ moderate, $r \geq 0.5$ large) and $\eta^2$ ($\eta^2 < 0.01$ no effect, $0.01 \leq \eta^2 < 0.06$ small, $0.06 \leq \eta^2 < 0.14$ moderate, $\eta^2 \geq 0.14$ large). Table 4 also presents both effect sizes.

| Metric | Exercise | Control | | Experimental | | p-value | U | r | $\eta^2$ |
|---|---|---|---|---|---|---|---|---|---|
| | | Mean | Median | Mean | Median | | | | |
| *Number of Test Suite Executions* | Olympic race | 73.76 | 46.00 | 105.30 | 88.00 | <0.001 | 9257.00 | 0.317 | 0.101 |
| | Exercise 1 | 11.04 | 6.5 | 26.28 | 16.00 | <0.001 | 10676.00 | 0.498 | 0.249 |
| | Exercise 2 | 18.42 | 11.5 | 37.94 | 25.00 | <0.001 | 10356.50 | 0.457 | 0.210 |
| | Exercise 3 | 24.16 | 5.00 | 22.64 | 18.00 | <0.001 | 8605.00 | 0.235 | 0.056 |
| | Exercise 4 | 20.14 | 7.5 | 18.44 | 7.00 | 0.105 | 7581.50 | 0.106 | 0.011 |
| *Active Time* | Olympic race | 7.55 | 8.00 | 11.39 | 11.00 | <0.001 | 12364.00 | 0.716 | 0.515 |
| | Exercise 1 | 1.89 | 2.00 | 5.45 | 5.00 | <0.001 | 13500.00 | 0.882 | 0.781 |
| | Exercise 2 | 1.87 | 2.00 | 2.14 | 2.00 | 0.001 | 8222.50 | 0.212 | 0.045 |
| | Exercise 3 | 1.89 | 2.00 | 2.18 | 2.00 | 0.002 | 8191.50 | 0.201 | 0.040 |
| | Exercise 4 | 1.90 | 2.00 | 1.61 | 2.00 | 0.008 | 5495.00 | 0.174 | 0.031 |
| *Participation Rate* | Olympic race | 0.47 | 0.00 | 0.58 | 1.00 | 0.102 | 7477.50 | 0.107 | 0.011 |
| | Exercise 1 | 0.75 | 1.00 | 0.86 | 1.00 | 0.034 | 7487.50 | 0.138 | 0.019 |
| | Exercise 2 | 0.75 | 1.00 | 0.90 | 1.00 | 0.002 | 7787.50 | 0.206 | 0.043 |
| | Exercise 3 | 0.68 | 1.00 | 0.85 | 1.00 | 0.002 | 7606.00 | 0.202 | 0.041 |
| | Exercise 4 | 0.74 | 1.00 | 0.73 | 1.00 | 0.865 | 5583.00 | 0.012 | 0.000 |
| *Dropout Rate* | Olympic race | 0.09 | 0.00 | 0.08 | 0.00 | 0.817 | 6692.50 | 0.015 | 0.000 |
| | Exercise 1 | 0.00 | 0.00 | 0.00 | 0.00 | 1.000 | 6750.00 | 0.000 | 0.000 |
| | Exercise 2 | 0.00 | 0.00 | 0.00 | 0.00 | 1.000 | 6750.00 | 0.000 | 0.000 |
| | Exercise 3 | 0.03 | 0.00 | 0.01 | 0.00 | 0.186 | 6597.50 | 0.086 | 0.007 |
| | Exercise 4 | 0.06 | 0.00 | 0.07 | 0.00 | 0.707 | 6582.00 | 0.025 | 0.001 |

Table 4. Results obtained for the metrics *Number of Test Suite Executions*, *Active Time*, *Participation Rate* and *Dropout Rate* for the Olympic race and each individual exercise: mean and median of the control and experimental groups, p-value and U obtained by the Mann-Whitney U test, effect sizes r and $\eta^2$.

In the Olympic race, the p-value obtained by the analysis for the *Number of Test Suite Executions* and the *Active Time* is smaller than α, whereas for the *Participation Rate* and the *Dropout Rate* is greater than α. Therefore, it can be assumed that there is significant difference between the control and experimental groups in the *Number of Test Suite Executions* and the *Active Time*, and according to r and $\eta^2$ the effect sizes are moderate and large, respectively. In addition, for these two metrics, the mean of the experimental group is greater than the mean of the control group. Regarding the *Participation Rate* and the *Dropout Rate¸* no significant differences are found; however, the mean of the *Participation Rate* is greater in the experimental group and the mean of the *Dropout Rate* is greater in the control group. So, the students in the experimental group executed more test suites, worked more days on the test improvement activity, worked more on all exercises and dropped out of the seminar exercises less than the students in the control group.

Analyzing each individual exercise, it can be observed that the results for the *Number of Test Suite Executions* and the *Active Time* are similar to the ones obtained in the Olympic race. In general, significant differences are found and the mean of the experimental group is greater than the mean of the control group. However, in the last exercise, the mean of the control group for both metrics is slightly higher (there is no significant difference for the *Number of Test Suite Executions*, whereas significant difference is found for the *Active Time*).

Regarding the *Participation Rate,* there is a significant difference between both groups in the first three exercises, with a small size effect, and the mean of the experimental group is greater than the mean of the control group. In the last exercise, there is no significant difference and the mean of the experimental group is slightly lower. On the other hand, no significant differences are found in any exercise for the *Dropout Rate.* The students in both groups started to drop out of the seminar exercises in the middle of the semester, which corresponds with exercise 3 (3% in the control group and 1% in the experimental group). In the last exercise, the dropout was slightly higher (6% in the control group and 7% in the experimental group).

Therefore, our findings from the Olympic race are in line with the first three exercises: the students in the experimental group worked more on the exercises and dropped out of them less than the control group. Only in the last exercise the control group seemed to be more engaged; however, no significant differences are found in three out of four metrics.

Overall, the null hypothesis is rejected in favor of the gamification experience when the engagement is measured with the four metrics. So, the engagement of the students who perform gamified software testing activities is higher than the ones who perform them in a non-gamified environment.

## 4.2 RQ2: Student performance

RQ2. Is the performance of the software testing students who carry out gamified activities higher than the ones who carry out them in a non-gamified environment?

To answer the research question RQ2, we state the null hypothesis of no difference in the performance of the software testing students who carry out gamified activities in comparison with the students who carry them out in a non-gamified environment. To test this hypothesis, we analyze the metrics *Effectiveness* and *Effectiveness Increase*. Again, we carried out the Mann-Whitney U test for independent samples to verify the null hypothesis of median equality with α=0.05, because we cannot assume the normality of the distributions (p-values < 0.001 was obtained in each Kolmogorov-Smirnov test). Table 5 shows the results for both metrics, considering the Olympic race and each exercise individually: mean and median for the control and experimental groups, p-value and U obtained by the Mann-Whitney U test. The effect sizes $r$ and $\eta^2$ are also presented in Table 5.

| Metric | Exercise | Control Mean | Control Median | Experimental Mean | Experimental Median | p-value | U | r | $\eta^2$ |
|---|---|---|---|---|---|---|---|---|---|
| *Effectiveness* | Olympic race | 81.35 | 93 | 88.43 | 98 | 0.008 | 8064.00 | 0.17 | 0.030 |
| | Exercise 1 | 90.64 | 100 | 93.79 | 100 | 0.086 | 7328.50 | 0.11 | 0.013 |
| | Exercise 2 | 83.35 | 100 | 92.30 | 100 | <0.001 | 8124.00 | 0.22 | 0.048 |
| | Exercise 3 | 78.75 | 100 | 86.95 | 100 | 0.003 | 7756.00 | 0.19 | 0.037 |
| | Exercise 4 | 78.05 | 100 | 82.64 | 100 | 0.300 | 6110.50 | 0.05 | 0.002 |
| *Effectiveness Increase* | Olympic race | 0.68 | 0.80 | 0.81 | 0.97 | 0.006 | 8119.50 | 0.18 | 0.031 |
| | Exercise 1 | 0.82 | 1.00 | 0.89 | 1.00 | 0.98 | 7301.50 | 0.11 | 0.012 |
| | Exercise 2 | 0.70 | 1.00 | 0.86 | 1.00 | <0.001 | 8163.50 | 0.23 | 0.051 |
| | Exercise 3 | 0.65 | 1.00 | 0.81 | 1.00 | 0.002 | 7839.00 | 0.21 | 0.043 |
| | Exercise 4 | 0.69 | 1.00 | 0.71 | 1.00 | 0.491 | 5966.00 | 0.05 | 0.002 |

Table 5. Results obtained for the metrics *Effectiveness* and *Effectiveness Increase* for the Olympic race and each individual exercise: mean and median of the control and experimental groups, p-value and U obtained by the Mann-Whitney U test, effect sizes r and $\eta^2$.

For both metrics, the p-value obtained for the Olympic race is smaller than α, so once again, it can be assumed that there is significant difference between the control and experimental groups, although the effect size is small. Moreover, the mean of the *Effectiveness*, as well as the mean of the *Effectiveness Increase*, are higher in the experimental group. Therefore, the students in the experimental group achieved better performance and they worked harder in the test improvement activity to increase the effectiveness.

The analysis of each individual exercise also reveals that the experimental group performs better: the difference is significant in both metrics in exercises 2 and 3, with small size effect, and the mean of both metrics is higher in the experimental group in all exercises. Despite the benefits of the gamification experience to improve both the *Effectiveness* and the *Effectiveness Increase*, we can observe a downward trend, mainly in the last two exercises.

Therefore, the null hypothesis is also rejected in favor of the gamification experience when the performance is measured with both *Effectiveness* and *Effectiveness Increase*. So, the performance of the students who carried out gamified software testing activities is higher than the ones who carried them out in a non-gamified environment.

## 5 Discussion

The main challenge we addressed in the present study was to measure and evaluate the impact of gamification on student engagement and performance. To deal with this challenge, we defined the metrics presented in Section 3.5 and analyzed the results obtained in the controlled experiment that was carried out. The results show that, overall, gamification contributes to the improvement of the students' engagement and performance. These findings are in line with those of most works that reported practical gamification experiences in software testing education. (Buckley & Clarke, 2018; Lőrincz et al., 2021; Sherif et al., 2020) compared the results achieved by gamified and non-gamified groups; (Lőrincz et al., 2021; Sherif et al., 2020) reported that the engagement was higher in the gamified one and (Buckley & Clarke, 2018; Sherif et al., 2020) also stated that the gamified group performed better. (Clarke et al., 2017; Costa & Oliveira, 2020; Elgrably & Oliveira, 2018; Fraser et al., 2019; Fu & Clarke, 2016; Sheth et al., 2015, 2013; Valle et al., 2017) did not make a comparison with a control group, but they observed that gamification had a positive impact on the student performance. In addition, the authors pointed out that the students evaluated the gamification experiences as very positive and they were engaged. (Clarke et al., 2019, 2022) evaluated whether increasing the use of a set of learning and engagement strategies that include gamification, referred to as LESs, improved the student performance. The authors compared the performance of the students minimally exposed to LESs with the performance of the students fully exposed to LESs and reported that the latter performed better. Additionally, (Clarke et al., 2020) evaluated the students' satisfaction when LESs were used, compared to the approach used in a previous course of their degree, and observed that the students were more satisfied when LESs were used. On the contrary, (Khandelwal et al., 2017), after comparing gamified and non-gamified groups, concluded that there was no impact of gamification on the student performance, despite most of the students being in favor of gamification. Similarly, (Jesus et al., 2020) stated that there was no difference in performance between gamified and non-gamified groups, and non-gamified groups were more engaged despite the fact that gamification attracted more students' attention. Both works reported short time length experiences and emphasized that more investigation is needed. In addition, (Jesus et al., 2020) indicated that the short duration of their gamification experience was a limitation of their research and a more real scenario would be an entire academic semester, leading to improving engagement and performance in the long-term. Those beliefs are consistent with our findings.

We have also found that in the first exercises the level of engagement of the students in the experimental group was higher than that of the students in the control group. In other words, their dropout rates were lower and both the interaction with SQLTest and the participation in all activities of the exercises were higher. In addition, we have noted that this engagement follows a downward trend in the last exercise. Regarding the student performance, the experimental group obtained better results in every exercise, and it also follows a downward trend in both experimental and control groups.

As a consequence of the downward trend observed, we faced a second challenge: analyzing whether the gamification impact remained constant over the whole experience or it varied in certain moments. To deal with this challenge, we analyzed the results obtained for the experimental group in each individual exercise.

The students in the experimental group spent more days working on the first exercise (the active time is twice as many days as in the second exercise), but they executed more test suites and participated more in the second exercise (44% more test suites executions and 5% more participation rate than in the first exercise). In addition, no student dropped out of the seminar in the first two exercises. The active time of the third exercise was similar to the second one, but the number of test suite executions and the participation rate decreased to values slightly lower than those of the first exercise (14% less test suites executions and 1% less participation rate than in the first exercise). Moreover, the dropout rate increased slightly in the third exercise to 1%. In the last exercise, the engagement decreased a bit more. The students worked half a day less in this exercise than in the second and third ones. Furthermore, the number of test suites executions and the participation rate were 18% and 14% lower, respectively, in the last exercise than in the third one. In addition, the dropout rate was slightly higher in the last exercise (7%). Therefore, the impact of gamification on the student engagement did not remain constant over the whole experience, as students seemed to be more engaged in the second exercise and slightly less engaged in the last one.

Regarding their performance, the effectiveness and the effectiveness increase were similar in the first two exercises (just slightly higher in the first exercise). In the third exercise, both the effectiveness and the effectiveness increase were 6%, approximately, lower than in the second exercise. Once again, in the last exercise, both metrics were lower than in the previous exercise. The effectiveness decreased another 5%, whereas the effectiveness increase was reduced by 12%. Therefore, the impact of gamification on the student performance varied slightly toward the last stages of the experience. Our findings do not seem to be in line with the results of (Fraser et al., 2019), which indicated that the students' performance improved throughout the semester. However, in their approach there was not a gamification experience that involved the whole semester, but the laboratory sessions were gamified independently through several games. As a result, the effects of a long gamification experience was not analyzed in that work.

In view of the results obtained in the analysis of the second challenge, which revealed that the impact of gamification was slightly lower in the last exercises, we addressed a third challenge: analyzing whether there was a link between gamification and both extrinsic and intrinsic motivation. To respond to this challenge, we analyzed the behavior of the students, mainly in the last exercise.

The slight reduction we observed may be caused by the gamification itself and the lack of benefit perceived by the students at the end of the race. In every exercise, some students worked to achieve 100% of test suite effectiveness in order to obtain the maximum 1,000 distance units given by that exercise, as well as the divine points they used to buy distance increments. These distance increments allowed them to advance faster in the race as they could accumulate more than 1,000 distance units per exercise. Other students did not work to achieve 100% of effectiveness, because they were given enough divine points to buy the distance increments for obtaining at least 1,000 distance units in that exercise.

This behavior is more noticeable in the last exercise. So, we observed that the students were more extrinsically motivated by the gamification experience to increase their performance in the first exercises because they received the benefits to advance quickly in the race. Consequently, almost 25% of the students had already achieved the finish line by the time the test improvement activity of the last exercise started.

When we analyzed in depth the behavior of the students in the last exercise, we observed that 8% of them did not carry out the test improvement activity, that is the last activity, because they had already reached the finish line, whereas 19% of the students did not participate in this activity maybe because their positions in the Olympic race ranking were too far away from rewarded positions. We also observed that 40% of the students reached the finish line by working on the test improvement activity, but 15% of the students did not complete this activity. Maybe they gave up because it was already impossible for them to reach the rewarded positions. So, some students dropped out of the test improvement activity of the last exercise because they had already reached the finish line and no more gamification benefits could be received, while other students gave up maybe because they were disengaged by the lack of expectations of reaching the rewarded positions. Consequently, this study cannot prove that the students were, finally, intrinsically motivated to carry out all academic activities of the seminars.

## 6   Threats to validity

We have identified several threats to validity in the current study, which are classified into four categories: internal, external, construct and conclusion validity.

*Internal validity*: The identified threats are as follows:

- Professor influence. Students may perform differently, depending on the professor that conducts the classes (Micari & Pazos, 2012). In order to avoid the professor influence in the results, the same professor has conducted all seminar sessions in both the control and experimental groups.
- Data collection and activity monitoring. For implementing the experiment, demographic data were collected and the game activity of the students was monitored. However, students could be reluctant not only to supply these data, but to be monitored constantly. Toward mitigating the student's reluctancy to participate in the gamified experience, the students were informed about the data to be collected and their treatment. Besides, the participation was voluntary and the students accepted the tools terms of use.
- Material to be used in class. This threat concerns the influence of the exercise domain and complexity, as well as the testing techniques to be used in the exercises, in the student's motivation. If the domain is not appealing enough or the program to be tested is not complex enough, students could perceive that creating test suites is boring or irrelevant. On the contrary, if the program is too complex with regard to the testing techniques taught in class, students could perceive that they do not have enough knowledge yet to deal with it. To mitigate this threat, each exercise deals with a different domain and the complexity of each exercise was in line with the concepts and testing techniques taught in the lecture classes. Besides, this complexity was increased progressively in each exercise as students acquired more knowledge.

- Injected defects. Some of the injected defects are more difficult to detect than others, so students may perform differently according to the difficulty level of the defects they have to detect. To mitigate this threat, all students in both the experimental and control groups have to detect the same defects in each exercise.

*External validity*: The identified threats are as follows:

- Student knowledge and skills acquired in lectures. If the students acquire different knowledge in the lectures in the experimental and control groups before carrying out the seminar exercises, the generalization of the results can be threatened. To mitigate this threat, the knowledge all students need to carry out the seminar exercises is taught in the lecture classes, using the same materials and methods in both experimental and control groups. Besides, the same professor conducted the lecture classes in both groups.

- Student knowledge and skills before enrolling in the course. If the students have different previous knowledge in the experimental and control groups or some of them have been trained in testing skills because they are already working on areas related to software development or software testing, again the generalization of the results can be threatened. To mitigate this threat, the knowledge all students need is taught in the lectures classes, as we stated above. Besides, the testing process and the testing techniques are only addressed in this degree course. In addition, the previous knowledge of the students in both groups was similar: the percentage of the students enrolled in the course more than once is quite similar (13% in the control group and 16% in the experimental group) and, based on the conversations between the professor and the students, only a few students start working in the meantime before they are enrolled in the course every year.

- Program representativeness: If the programs to be tested lack complexity, they could not be representative enough of industrial practice. To mitigate this threat, the programs are based on real-life applications.

*Construct validity*: The threat concerns the metrics used to answer the research questions. The use of metrics that do not describe the student's engagement and performance could produce misleading results. In order to mitigate this threat, we used student's participation metrics to measure the student's engagement and test suite effectiveness metrics to measure the student's performance. Both student's participation and test suites effectiveness are widely accepted metrics in the literature (Fredricks et al., 2004; Papadakis et al., 2019; Ruiperez-Valiente et al., 2021).

*Conclusion validity*: The threat concerns the researcher conclusions. To avoid incorrect researcher's interpretations of the results obtained, we carried out statistical analyses for each research question. We utilized the statistical test and effect sizes generally used when the normality of a distribution cannot be assumed.

## 7 Conclusions

This work presents a long gamification experience that was designed and conducted to motivate the students to create effective test suites, and makes a comparison with a non-gamified experience. The results show that the gamification benefits the improvement of both student engagement and performance. We have also observed a slight reduction of the engagement in some students toward the last stage of the experience, when they perceived that no more rewards were going to be received by keeping on working on the gamification experience. The statistical analysis indicates that the differences are significant.

In addition, the study confirms how the design of the gamification strategies is crucial for engaging students. The results obtained show that the rewards in this gamification experience acted as powerful extrinsic motivators that kept the students motivated until they perceived they were not going to get more rewards for their work. Arguably, because of the particular gamification design of this experience, the link between gamification and intrinsic motivation could not be proved. Nevertheless, the results did prove that gamification succeeds in keeping extrinsic motivation and improving performance for long periods of time. While designing a gamification experience, the professors should distribute the motivating stimulus throughout the whole experience, so that the engagement lasts until the end.

As a part of our future work, we will readjust the design of the gamification experience to include new rewards that increase student engagement in the last seminar exercises, so that we can overcome the challenge of keeping the engagement from decreasing slightly toward the last stage of the experience. In addition, we will study the effects of gamification in our software testing course during several academic years as part of a longitudinal study.

**Acknowledgements**

This work was supported by projects PID2019-105455GB-C32 and PID2019-105455GB-C33 funded by MCIN/AEI/10.13039/501100011033 (Spain).